# Reliability Disasters: a follow up

## by

## TP Davis[1]

The recent article in the August 2020 issue of *Quality Progress* "Reliability Disasters" by Doganaksoy Meeker and Hahn (henceforth DMH) raises some important questions about engineering for reliability, in particular the challenge referred to in their opening sentence, that "the elapsed time between when the product is designed and built and when reliability information is forthcoming", the implication being that this elapsed time is too long. The purpose of this follow up article is to show how a pro-active approach to reliability can be realized with an innovative use of the design FMEA (Failure Mode and Effects Analysis)[2] or dFMEA, which can then help to reduce the elapsed time between when an engineering prototype is available for reliability evaluation. Additionally, the FMEA can accommodate many of the reliability tools & methods advocated by DMH such as identifying key failures modes early in development, Accelerated Life Tests and up-front experimentation, physical models and so-on. As DMH say later in the article "The problem is surely not with the concept of proactive reliability but the manner in which it has been applied".

The primary focus of the approach to reliability engineering outlined in this article is to use the FMEA to choose the design that fails the least rather than trying to predict how often the chosen design will fail[3]. In other words, we change the paradigm within which the reliability engineering is done. We aim to show that this paradigm shift is necessary for engineers to take control of the elapsed time highlighted by DMH.

Engineering is about disturbing the current state of nature to create something tomorrow that does not exist today. The rules that describe how nature works is of course the domain of *science,* while *engineering* uses these rules to create new entities which do not occur naturally. The connection between science and engineering is nicely encapsulated in the definition of engineering provided by ABET[4]. Designs have to be developed according to the rules determined by Mother Nature, and although for the most part we understand the rules (e.g. the rules of geometry, the laws of motion and the properties of materials), engineering entities involve many interfaces between interacting components which means that we can't always easily tell ahead of time how things will work out. This uncertainty provides the conditions for potential failure modes to be overlooked and for their effects to propagate.

Therefore, the consequences of doing engineering could be thought of as progressing through the following three stages: -

---

[1] Tim Davis was latterly the Henry Ford Technical Fellow for Quality Engineering at the Ford Motor Company. He led Ford's technical team that investigated and resolved the Firestone tire crisis. He is currently senior statistician with We Predict Ltd. In the UK, as well as running his own quality engineering consultancy. He is a senior member of ASQ.

[2] In this paper we concentrate on the *design* FMEA because this is the form of the FMEA that is best suited to tackling the reliability issues highlighted in DMH.

[3] That is not to say that field reliability predictions are not useful – for example to plan warehouse capacity for spare parts and after-market service capacity, but in this author's experience predictions to the nearest order of magnitude are usually sufficient.

[4] The Accreditation Board for Engineering and Technology (ABET) connects engineering with science with the following definition "engineering is the profession in which a knowledge of the mathematical and natural sciences, gained by study, experience, and practice, is applied with judgment to develop ways to utilize, economically, the materials and forces of nature for the benefit of mankind".

1. Simply by doing engineering the conditions for failure modes[5] to emerge and flourish is created (not intentional, but inevitable).
2. These potential failure modes must be sought out and found (or detected[6])
3. A countermeasure for the failure mode to prevent[7] it from occurring, must be developed, verified that it works, and deployed into the product design.

Once the engineering project is underway the only variable under the control of the engineering team is when they choose to execute stage 2. If this stage is executed as soon as possible after stage 1., there will be more time to develop a countermeasure by the time that the team get to stage 3. This alludes to the elapsed time dilemma mentioned by DMH.

We begin by discussing these stages within the framework of a typical Product Development Process (PDP) together with a discussion of how failure modes manifest themselves therein.

**The Product Development Process[8]**

The Product Development Process governs the sequence of work activity from initial program approval (the Mission) through to delivering the final design to its field of operation.

Figure 1 illustrates a typical PDP, through a sequence of phases that progress in time. We note that the PDP synchronizes the flow of *information* (i.e. data which represents the current knowledge about the design based on tests and evaluations of transient artefacts such as prototypes and computer models) with updates to *material* (the specification of the current prototype).

---

[5] A failure mode in its broadest sense is an event that degrades the utility of the design, and will require a mitigation action (preferably a countermeasure i.e. a change to the design) to restore utility
[6] The word *detect* is introduced here to emphasise the connection to the FMEA (discussed later)
[7] "Detect and Prevent" is the proactive version of "find and fix".
[8] For more detail on Product Development as a process, see *Product Design and Development* by KT Ulrich & SD Eppinger (3rd edition, 2004), published by McGraw Hill, in particular Chapter 2.

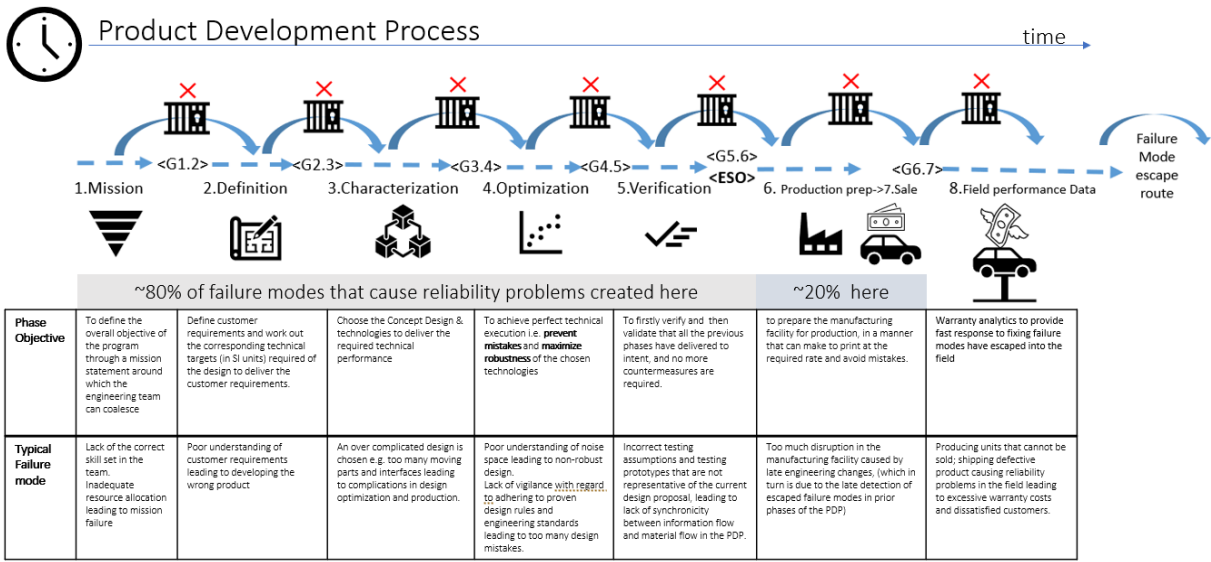

**Figure 1** A schematic of the PDP with reference to automotive engineering. (the PDP illustrated here can be easily adapted to the development and sale of other commercially available products). The workflow in each phase is illustrated by - - ->. It is conjectured that each phase of the PDP can generate failure modes which if not corrected with a countermeasure in the same phase, the failure mode will "escape" into the next phase whereby the opportunity to deploy the most appropriate countermeasure will be lost (see text). The primary objective of using the FMEA in the PDP is to detect failure modes as they are created so that a countermeasure can be deployed prior to moving to the next phase, which should be the primary focus of the gateway reviews, denoted <Gx.y>. In this way the escape path for the failure mode is cut off and ultimately by chain reaction field reliability is improved. **Note:** The gateway designated <G4.5>, which is the transition from the optimization phase to the verification phase is sometimes referred to the "engineering sign-off gateway" or <ESO>.

A key idea in the PDP is to ensure that the flow of information is *synchronous* with the updating of material (i.e. ongoing modification to prototypes) as the design is improved during development. That is, as knowledge is improved, prototypes are updated to reflect the current knowledge of the design established during the work in the preceding phase.

Note that within the PDP defined in Figure 1, we can conclude that it is required to take the countermeasure for the failure mode *in the phase that the failure mode was created*. For example if an incorrect understanding of customer requirements is made in the definition phase (phase 2 in Figure 1) no amount of subsequent design characterization in phase 3 or optimization in phase 4 will correct this error, so it needs to be discovered prior to proceeding through Gateway <G2.3>. Thinking about failure modes in this manner illustrates that a failure mode that escapes into the field (which if the effect is severe enough creates a "reliability disaster") is just a special case of a failure mode escaping past a gateway in the PDP and into the next phase. We will see shortly how the FMEA can help to identify potential escape points for failure modes. Preventing failure modes from escaping past their creation phase by detecting them quickly and deploying a countermeasure in time for the next gateway is known as ***failure mode avoidance***.

The four reliability examples cited by DMH (discussed later) have two themes in common, namely ***i)*** that clear communication between the technical engineers and management[9] regarding the level of engineering knowledge is vital to avoid any misunderstanding of attendant risks to the current plan, and ***ii)*** determining what went wrong ***in hindsight*** is relatively straightforward[10]. Who could forget the experiment that the physicist Richard Feynman conducted during the Senate Committee hearings investigating the Challenger disaster to demonstrate that the O-ring material in the Solid

---

[9] By management, we mean members of the organization, who will assume the greater responsibility should things go wrong.

[10] There are several Problem-solving algorithms to assist engineering teams in determining the (root) cause of a failure mode, for example the 8D problem solving method promoted by the Ford Motor Company, and the practical problem solving (PPS) approach of the Toyota Motor Corporation.

Rocket Boosters ('SRB's) lost its compliance at low temperatures[11]? One can only believe (hope) that prior to the launch of the shuttle had a suitably calibrated and instrumented version of Feynman's experiment been conducted by NASA or Morton Thiokol engineers and the results communicated to management, then the launch of the Challenger would have been delayed as a mitigation action and a re-design of the seal and joint then commissioned to improve the robustness of future launches to external temperature[12]. So, a critical reliability question in the context of the DMH article is "how can we identify all potential failure modes and evaluate required countermeasures prior to exposing the product to its field conditions"?

In other words, "how can we turn hindsight into foresight"?

The examples cited by DMH can all be seen initially as *robustness* problems – that is the proposed engineering design turned out to be sensitive to encountered environmental conditions (which in the robustness context are a special case of what are referred to collectively as *noise factors[13]*).

In the battery example the compounding of the epoxy, the lead composition, and physical design of the battery combined to create a noise condition, which caused blister corrosion leading to seal leakage around the cathode[14]. In the refrigerator compressor example, the noise was a degradation of material properties[15] caused by excessive running temperature. In the Challenger example the key noise factor was external temperature which in turn reduced the temperature of the O-ring material, and in the Firestone tire example, the key noises were tire pressure and load which when combined with high ambient temperature, vehicle speed and tire age generated excessive internal heat between critical components in the tire leading to separation of the tire tread from the steel belt (tread to belt separation - TBS) [16].

Of course, to be pro-active, in a manner advocated by DMH one must be cognizant of the set of noise factors that cause robustness problems during the PDP so that countermeasures that make the design robust can be incorporated into the program prior to the engineering sign-off event[17]. Once these robustness lessons have been learned, they can be incorporated into updated design rules and engineering standards (forming part of a *design catalogue* which reflects emergent practice) so that any subsequent occurrence of the failure modes can be regarded as a *mistake* due to not adhering to the design rules within the catalogue. Hence, we have two high level causes of reliability failures, **1) *robustness problems*** (sensitivity of the design to noise factors) and **2)**

---

[11] A video of Feynman doing this experiment can be found on at https://www.youtube.com/watch?v=raMmRKGkGD4
[12] Which is what happened during the 2-year hiatus between Challenger (STS-51-L) and the next launch (STS 26 "return to flight"). The redesigned joint was stiffened, to limit flexibility and eliminate potential leak paths. There is of course a more fundamental design question which could have been asked during the design characterization phase of Figure 1a - "Why did the SRB have so many joints?" – we do not discuss that design question here
[13] Noise factors are sources of uncontrollable variation experienced by the product in the field, which impacts on the ability of a design to function. There are five general categories of noise factor **1.** Variation part-to part in characteristics (e.g. dimensions) caused by manufacturing at rate**, 2**. Variation in part characteristics (e.g. wear, degradation) caused by exposure to repetitive demand, **3.** Variation in customer usage and duty cycles caused by variety of end-users, **4.** Variation in external environment due to differences in climate and infra-structure conditions across markets, and **5.** Variation in the Internal (to the system) environment caused by interactions between adjacent components in the system. Note that "environmental conditions" is only one of the five categories of noise (noise 4), but all 5 sources of noise should be considered in reliability analysis
[14] A detailed account of this case is contained in Cannone, A. *et al*. "The round cell: promises vs. results 30 years later." *INTELEC 2004. 26th Annual International Telecommunications Energy Conference* (2004): 401-410.
[15] For piece-cost reasons two rotating parts were made with powdered metal steel, rather than hardened steel
[16] Note that the cause of these TBS failures is a complicated interaction of five distinct noise factors perhaps explaining in part why these failures escaped into the field undetected during the PDP with such devasting consequences.
[17] The "Engineering sign off event <ESO> should occur prior to the verification phase of the PDP (see Figures 1) Ideally there should be no design changes after <ESO> (<G4.5> in Figure 1). If there *are* changes after this point this will be due to detecting failure modes later (too late) in the PDP.

*mistakes*[18] (not implementing design features that are known a-priori to avoid the subject failure mode). We can think of mistakes as the absorbing state so that as robustness problems are learned and overcome, we update the design catalogue, which when followed, will prevent the failure mode from occurring again for the identified cause. In this way ongoing reliability improvement to evolving designs through the product life cycle is achieved (eventually) simply by avoiding mistakes.

We will refer to the DMH examples to illustrate how some of things we are recommending *could* perhaps have been pro-actively applied in these cases.

## Failure Mode and Effects Analysis (FMEA)

We assume a working knowledge of the FMEA from this point forward and only emphasize the salient points for failure mode avoidance, namely recognizing "escape points" for failure modes, and the development of effective detection controls that will excite the failure mode and that can be executed prior to the engineering sign-off gateway. We put primary emphasis on *detecting* the failure mode as soon as possible after it has been *created*, recognizing that the point in the PDP where the failure mode is created is usually not the same point in time as when it is *naturally observed*.

Figure 2 shows a facsimile of a typical FMEA document, which as read from column to column reflects the workflow in each phase of the PDP. Note the left most columns (Ideal function through to cause) generate Information, which eventually *transitions* into material (the Prevention Control or counter measure). The FMEA is then a "sub-routine" inside in the PDP. Therefore, it naturally lends itself to becoming a project management tool aimed at failure mode avoidance.

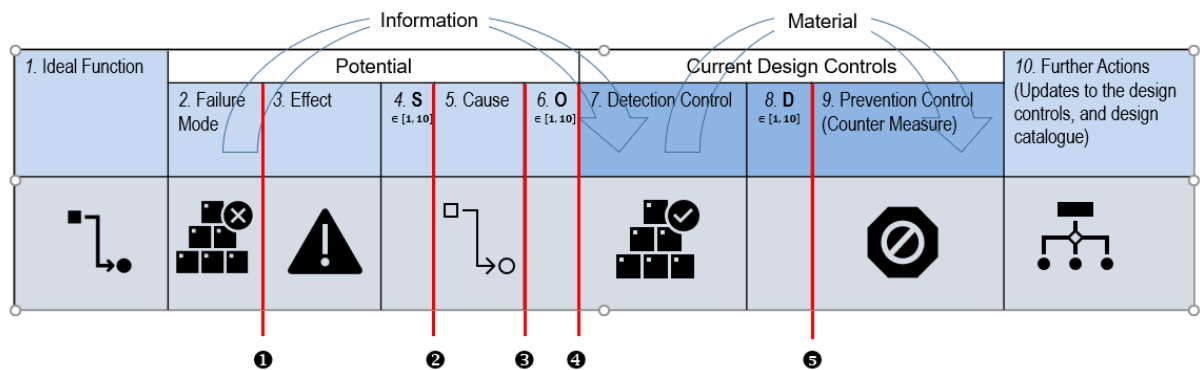

**Figure 2.** a schematic of a design FMEA, which involves a flow of information (the potential failure mode, its effect and cause) through to the generation of new material (the Prevention Control or countermeasure), via a Detection Control (an event that can excite the subject failure mode due to the identified cause). "Escape points" for failure modes are highlighted by the vertical red lines (see text for a discussion).

The FMEA can be used to ensure that material flow and information flow and remain synchronised, thus minimizing dis-order[19] in program delivery. As such the FMEA, like the PDP itself, is a special case of *value stream mapping*[20].

Failure modes escape if a step in the FMEA is not completed, which compromises all subsequent activity. Escape points ❶ thru ❹ in Figure 2 are concerned with correctly identifying the failure mode and cause and correctly assessing the severity of the effect and occurrence of the cause[21]. If such escapes happen the last line of defense is the Detection Control. This is the event (either a physical test or analytical evaluation) which is designed to excite the failure mode. To assess the efficacy of the detection control, the utility of the detection score (D) is crucial. A score[22] of 1 means that the chosen detection control is certain to excite the failure mode, whereas a score of 10 implies that the detection control does not excite the failure mode at all, with graduating scores in between[23]. If we underestimate the efficacy of the chosen detection control, then the failure mode will escape here (escape point ❺ in Figure 2), and then into the field of operation.

---

[18] Professor Don Clausing during his time at Xerox proposed these classifications for failure mode causes. See for example, "Improving System Reliability by Failure Mode Avoidance, including four Concept design strategies" by D Frey & D Clausing published in Volume 8 of the journal Systems Engineering in 2005.
[19] This dis-order is often manifested by fabricating prototypes that are "unrepresentative" in the sense that they do not represent the current knowledge as to which countermeasures are required.
[20] See the Wikipedia article `https://en.wikipedia.org/wiki/Value_stream_mapping`

The idea is to use the FMEA is to help reduce the elapsed time between when the product is designed and built to when reliability information is forthcoming. Here, though, the reliability information is generated from an effective detection control[24] (ideally executed within the phase that the failure mode was created) rather than field data.

Once the FMEA is complete[25] it takes the form illustrated in Figure 3.

With appropriate focus on failure mode escape points and effective and timely detection controls, the FMEA can turn hindsight into foresight and provide for a proactive approach to reliability, while providing a channel for clear communication between the design engineers and management.

**Figure 3**: A Schematic of how the FMEA develops as a program evolves. Note the bifurcation due to Ideal function possibly having more than one failure mode and failure modes having possibly more than one cause.

---

[21] The product of the severity and occurrence scores is called the criticality index
[22] Because the detection control is fundamental to Failure Mode Avoidance, we give a suite of detection scores here: 10 = cannot detect (p=0), OR no prevention control is proposed or planned for the subject failure mode; 9 = very remote chance to detect (p<0.01); 8 = remote (p<0.1); 6 = low chance (p<0.5); 5 = moderate chance (p>0.75); 3 = high chance (p>0.9); 2 = very high chance (p>.99); 1 = certain to detect (p=1). Note that the probabilities (p) are not frequencies (so a required sample size is not implied) but rather a degree of belief as to the effectiveness of the detection control to excite the failure mode due to the subject cause.
[23] which is equivalent to the absence of a countermeasure.
[24] Preferably a detection control with D≤2.
[25] We might argue the FMEA is never be complete, but classification systems for failure modes, causes, and noise factors can provide the engineering team with an indication as to whether anything might have been overlooked.

## Summary

How might the FMEAs for the four examples cited in DMH have looked if what had been learned in hindsight could have been pro-actively applied with foresight, with reference to the failure mode escape points?

The AT&T battery example

The escape point in this example was that the blister corrosion was not observed in the detection control (the ALT). Either *i)* this type of corrosion had not been identified as a potential failure mode, (escape point ❶ in Figure 2) or *ii)* it had and because the ALT did not excite it, it was assumed that the proposed round cell battery design was already immune to this failure (escape point ❺ in Figure 2).

The GE refrigerator compressor example

The escape point in this example was that the severity of an observed failure mode (the discoloration of material inside the compressor) was underestimated (escape point ❷ in Figure 2), so that during development insufficient attention was placed on integrating a countermeasure for this failure in the design.

The Firestone tire crisis[26]

Radial pneumatic tires are a mature design concept, and as such active detection of failure modes during vehicle development relies heavily on tires meeting the criteria laid out in Federal Motor Vehicle Safety Standard (FMVSS) tests protocols. These tests are aimed at *verifying* basic structural integrity of the tire at various speeds, rather than *detecting* new failure modes, and although TBS failures where a known phenomenon, the FMVSS tests did not excite this failure mode (escape point ❺ in Figure 2). In order to replace the Firestone tires in the market, the Ford Motor Company d*e*veloped, via a statistically designed experiment, a laboratory based rig test with the appropriate settings of the identified noise factors to excite TBS, so that replacement tires could be verified as being immune to TBS for the cause identified. This test was added to an updated FMEA as a detection control for TBS on future vehicle programs which required new tire development (thus plugging the escape point), along with updates to the design catalogue with particular regard to temperature generation within the tire across a range of tire pressures and loads.

---

[26] This narrative is based on this authors' personal involvement in resolving the Firestone tire crisis. For a partial account see Krivtsov VV, Tananko DE, and Davis TP. "A regression approach to tire reliability analysis", 2002, *Reliability Engineering & System Safety*, volume 78, number 3, pp 267-273.

The Challenger disaster

Following Richard Feynman's demonstration of the loss of seal compliance at temperatures at or below freezing, the FMEA for the developing the field joint seal in the SRB, could have looked something like that shown in Figure 4.

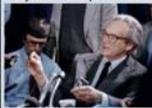

**Figure 4:** How an FMEA for the Challenger case might have looked if Feynman's experiment had been used as the detection control prior to launch and during development. Of course, the procedure would need to have been suitably instrumented and calibrated so that the performance of the seal could be accurately measured in the testing laboratory over the appropriate range of temperatures. An occurrence rating of 5 is given here because about half of the Shuttle launches prior to STS-51-L showed some damage to the field joint O-ring. [see Figure 1(b) in DMH]. Note that this example nicely illustrates that detection controls do not need to be executed on the complete system, but rather a conveniently selected sub-system, exposed to the correct noises.

## **Conclusion**

Failure Mode Avoidance utilizes the FMEA in ways that a statistical approach to reliability does not by putting a major emphasis on the timing and effectiveness of Detection Controls relative to the PDP. In fact, in this context it may be better to think of FMEA as Failure Mode and Effects *Avoidance* rather than failure mode and effects *analysis*.

Of course there is no guarantee that we can turn hindsight into foresight but a well-executed FMEA with emphasis on failure mode and cause identification and the use of detection controls well before the verification phase of the PDP puts the elapsed time to generate reliability information firmly in the hands of the engineer, which surely increases the chances of developing reliable designs.

By reliable, we mean choosing the design that fails the least, and better than that we cannot do.

As the examples in DMH illustrate, resources can always be found to fix a failure mode that has escaped into the field and caused a reliability disaster, but in the end, there is no point in waiting until one's back is against the wall before reading the writing upon it.